\documentclass[a4paper,twocolumn]{esapub2005} 
\usepackage{times}
\usepackage[numbers]{natbib}
\usepackage{graphicx}
\usepackage{amssymb}
\usepackage{amsmath}
\usepackage{subfigure}
\usepackage{wrapfig}

\title{INTEGRAL observations of HER X-1}
\author[1]{D. Klochkov}
\author[1]{R. Staubert}
\author[2]{S. Tsygankov}
\author[2]{A. Lutovinov}
\author[3]{K. P. Postnov}
\author[3]{N. I. Shakura}
\author[3]{S. A. Potanin}
\author[4]{C. Ferrigno}
\author[1]{I. Kreykenbohm}
\author[5]{J. Wilms}

\affil[1]{Institut f\"ur Astronomie und Astrophysik, University of T\"ubingen, Sand 1, 72076}
\affil[2]{Space Research Institute, Moscow, Russia}
\affil[3]{Sternberg Astronomical Institute, 119992, Moscow, Russia}
\affil[4]{INAF IASF-Pa, via U. La Malfa 153, 90146 Palermo, Italy}
\affil[5]{Department of Physics, University of Warwick, Coventry, CV8 1GA, UK}

\begin{document}

\keywords{X-ray binaries; accretion disks; neutron stars}

\maketitle

\begin{abstract}
First results of observations of the low mass X-ray
binary Her X-1/HZ Her performed by the INTEGRAL satellite in
July-August 2005 are presented. A significant part of one 35 day
main-on state was covered. The cyclotron line in the X-ray spectrum
is well observed and its position and shape, as well as its
variability with time and phase of the 1.24 s pulsation are explored.
X-ray pulse profiles for different energy bands are studied throughout
the observation. The pulse period is found to vary on short time
scales revealing a dynamical spin-up/spin-down behavior. Results of
simultaneous optical observations of HZ Her are also
discussed.
\end{abstract}

\section{Introduction}

Discovered in 1972 by the UHURU satellite \citep{Giacconi73,Tananbaum72}, Her X-1 is one of the most extensively studied accreting pulsars. Being part of a low mass X-ray binary it shows strong variability on very different time scales: the 1.24s spin period of the neutron star, the 1.7\,d binary period, the 35\,d period of precession of the warpted tilted accretion disk \citep{GerBoy76,Shakura99,Ketsaris00}, the 1.65\,d period of the preeclipse dips \citep{Giacconi73,Tananbaum72}. Due to the high orbital inclination of the system ($i>80^{\circ}$) the counter-orbitally precessing warped accretion disk around the neutron star blocks the observers view the X-ray source during a substantial part of the 35\,d period. This gives rise to the alternation of so-called {\em On} (high X-ray flux) and {\em Off} (low X-ray flux) states. 
The 35\,d cycle contains two {\em on} states -- the {\em main-on} and the {\em short-on} -- separated by $\sim7\div8$\,d {\em off} states. The X-ray flux in the middle of the main-on is $\sim4\div5$ times higher than that in the middle of the short-on. The sharp transition from the off-state to the main-on is called the {\em turn-on} of the source. Turn-ons are usually used for cycles counting.  The 35\,d period manifests itself also by the changing shape of the X-ray pulse profiles \citep{Soong90a, Truemper86, Deeter98, Scott00} and through the modulation of the optical light curve \citep{GerBoy76, HowarthWilson83}.

\begin{figure}
\resizebox{\hsize}{!}{\includegraphics{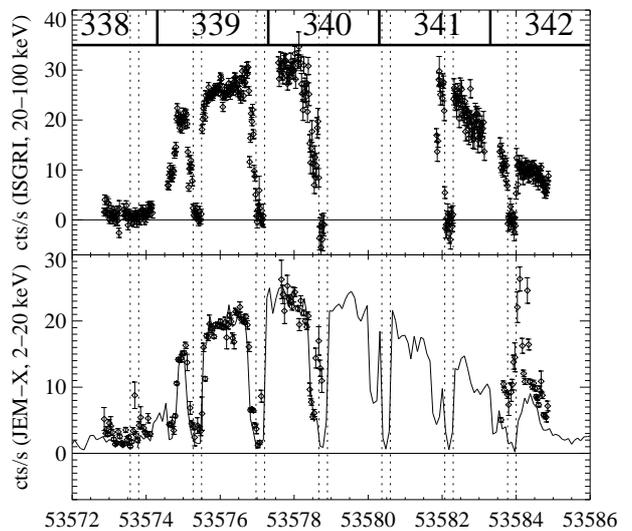}}
\caption{\footnotesize Low resolution light curves of Her X-1. Vertical dashed
  lines mark X-ray eclipses. On top of the upper panel revolution numbers are indicated.
  The solid curve in the bottom panel shows the ASM RXTE light curve averaged 
  over many 35\,d cycles.}
\label{lc}
\end{figure}

In spite of a large amount of observational data the system is still poorly understood. Physical mechanisms responsible for the disk precession, X-ray dips and evolution of X-ray pulse profiles are highly debated. In this work we present X-ray observations of Her X-1 performed by the INTEGRAL observatory.
High sensitivity complemented by very good spectral and timing resolution  
makes \hbox{INTEGRAL} exceptionally useful for performing observations which can lead to new insight into the physics which is at work in this system.

\section{Observations and data processing}

\begin{figure*}
\centering
\includegraphics[width=17cm]{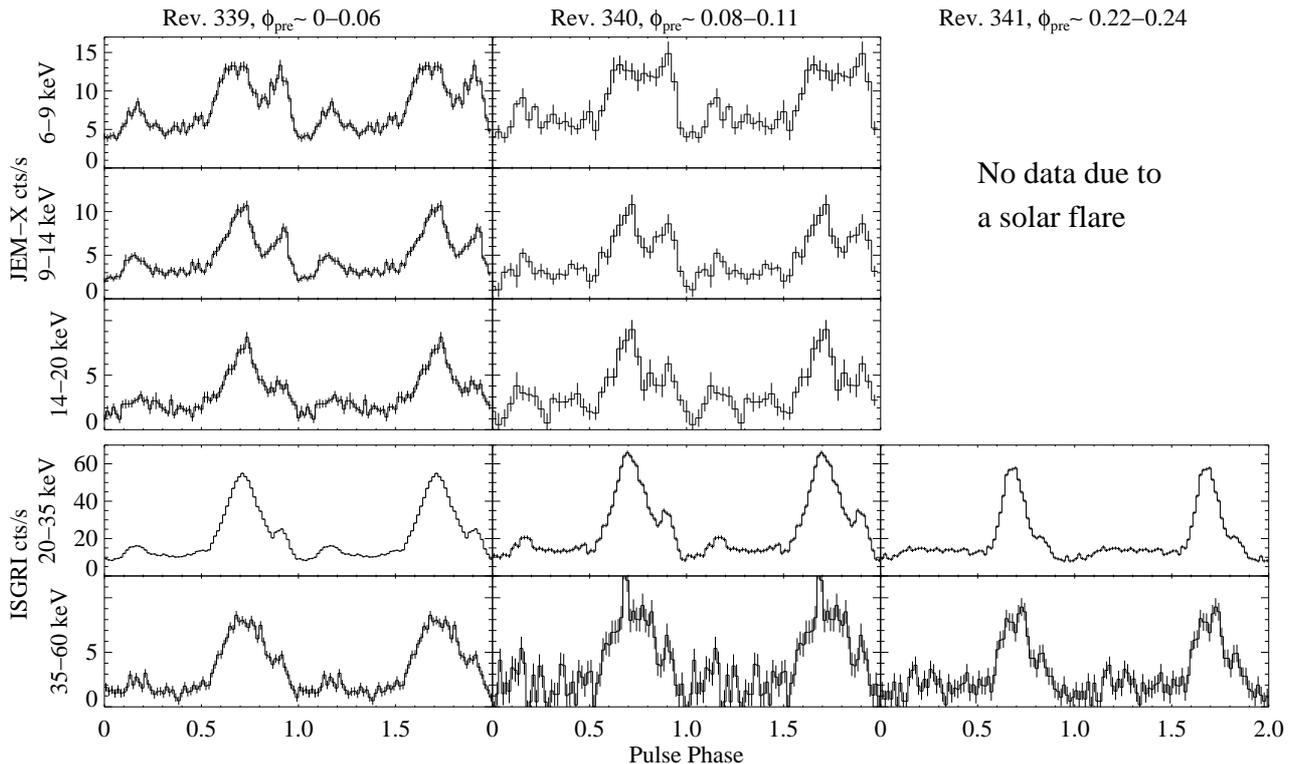}
\caption{\footnotesize X-ray pulse profiles of Her X-1 for different energy
    bands and time intervals.}
\label{pp}
\end{figure*}

The accreting X-ray pulsar Her X-1 was observed by \hbox{INTEGRAL}
on July 22 -- August 3, 2005 (MJD: 53573-53585).
The observations include $\sim 5$ orbital revolutions of the
satellite: revs. 338, 339, 340, 341 and 342. 
They partially cover a main-on state of the source: 35\,d phases $\phi_{\rm pre}\sim0-0.11$ and $\phi_{\rm pre}\sim0.20-0.28$ ($\phi_{\rm pre}=0$ is the phase of the turn-on of the source).
In our analysis we use the data
obtained with the instruments JEM-X ($\lesssim 20$\,keV) and ISGRI
\hbox{($\gtrsim 20$\,keV)}.
The data processing is performed with the version 5.1
Offline Science Analysis (OSA) software distributed by ISDC \citep{Courvoisier03}. 
Additional gain correction of the photon energy based on the position of the tungsten background line was done. A 1\% systematic error to both JEM-X and ISGRI data sets has been added \footnote[1]{http://isdc.unige.ch/index.cgi?Support+documents}.
For the pulse-phase-resolved spectroscopy we also used the software developed at IASF, Palermo \citep{Mineo_etal06}.
The spectral analysis is performed with the XSPEC package v. 11.3.
The results of the analysis of averaged spectra are independently checked using the software package developed at Space Research Institute, Moscow
\citep[see e.g.][]{Tsygankov06a}.
Figure~\ref{lc} shows low resolution light curves obtained with ISGRI (top)
and JEM-X (bottom). The turn-on occurred at MJD $\sim 53574.7$ corresponding to orbital phase $\phi_{\rm orb} \sim 0.7$. The solid line on top of the \hbox{JEM-X} light curve represents the ASM RXTE light curve 
averaged over {\em many} 35\,d cycles
and stretched vertically to match the one from JEM-X. 
Pre-eclipse dips (sharp flux decrease just before the eclipse) are clearly 
seen. All JEM-X data from revolution 341 are rejected because of a strong 
solar flare.

\section{Pulse profiles}

We extracted X-ray pulse profiles for different energy bands and time intervals. Only the data corresponding to the on-state of the source are presented (revs. 339, 340 and 341). We also excluded dips and X-ray eclipses.
During the off-state (revs. 338 and 342) the data are very noisy and show marginal sinelike pulsations. 
It is seen (Fig.~\ref{pp}) that pulse profiles change significantly in shape during the period of
observations. Their shape is typical for Her~X-1. Analogous pulse profiles were observed by RXTE at different times \citep[see e.g.][]{Kuster05}.

\begin{figure}
\resizebox{\hsize}{!}{\includegraphics{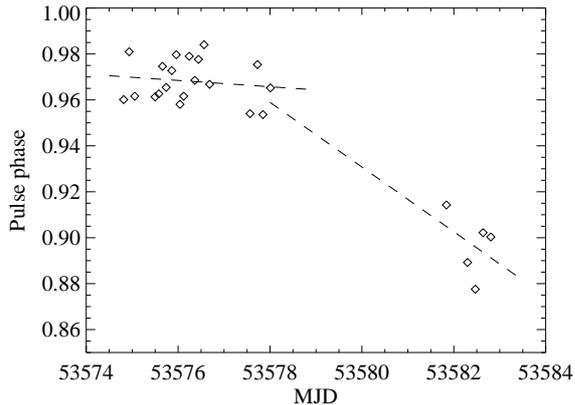}}
\caption{\footnotesize Pulse phase of the sharp decay after the
  trailing ``shoulder''  of the main peak used as a phase reference of the pulse period (see text).
  The apparent change of the
  slope (dashed lines) indicates a pulse period change by $\sim2\times10^{-7}$\,s
  on a time scale of few days.}
\label{pper}
\end{figure}

The value of the pulse period found with the standard epoch folding method is $1.237758(36)$\,s.
This period has been used for constructing the pulse profiles and performing the pulse-phase-resolved spectroscopy (see below). Due to limited precision of the epoch folding method it cannot be used to study the short-time-scale variations of the pulse period. Such variations can be explored by measuring the relative phase shift of corresponding details between different pulse profiles.
From the analysis of RXTE observations of Her X-1 it was found
that the most stable and best measured feature
of the pulse profile is the steep decay after
the trailing ``shoulder'' of the main peak near pulse phase 0.9.
The position of this decay does not depend on the shape or position of
peaks and varies with time much less than any other detail of the
pulse profile. It therefore provides a good phase reference.
Figure~\ref{pper} shows the pulse phase of the decay as a function of time. If the period of pulsations were constant the phase would be linear with some constant slope (with slope zero for the correct period). The apparent change of the slope (dashed lines in Fig.~\ref{pp}) indicates a change of the period by $2\times10^{-7}$\,s on a time scale of few days. 


\section{Averaged spectrum}

\begin{figure}
\resizebox{\hsize}{!}{\includegraphics[angle=-90]{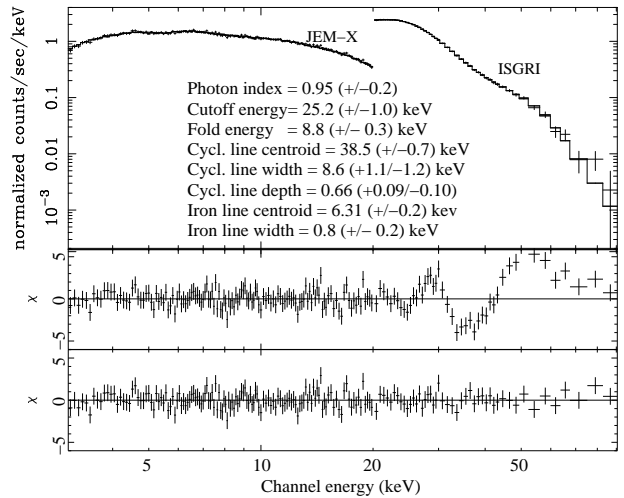}}
\caption{\footnotesize {\em Top panel}: the averaged X-ray spectrum of 
Her X-1 (best fit model parameters with $1\sigma$ uncertainties are shown); 
{\em middle panel}: residuals after fitting the spectrum without the cyclotron 
line; {\em bottom panel:} residuals after fitting the spectrum with the adopted model (see text).}
\label{spe}
\end{figure}

For constructing the average spectrum, we used the data from revolutions 339 and 340. During this interval the source is in the On state and both JEM-X and ISGRI data are available. As before, eclipses and dips are excluded. The resulting spectrum is modeled by a power law with an exponential cutoff (\textsf{highecut} in XSPEC). A gaussian emission line is added at $\sim 6.4$\,keV to account for the iron line. The cyclotron line is firmly observed and modeled by a gaussian absorption line. The averaged spectrum is shown in Fig.~\ref{spe} along with the best fit model parameters and $1\sigma$ uncertainties. $\chi^2_{\rm red}$ = 0.78/165 d.o.f. The position of the line ($38.5\pm0.7$keV) is consistent with that measured by RXTE at a similar time \citep{Staubert06c, Staubert06d}.

\section{Pulse-phase-resolved spectroscopy}

\begin{figure*}
\centering
\includegraphics[width=17cm]{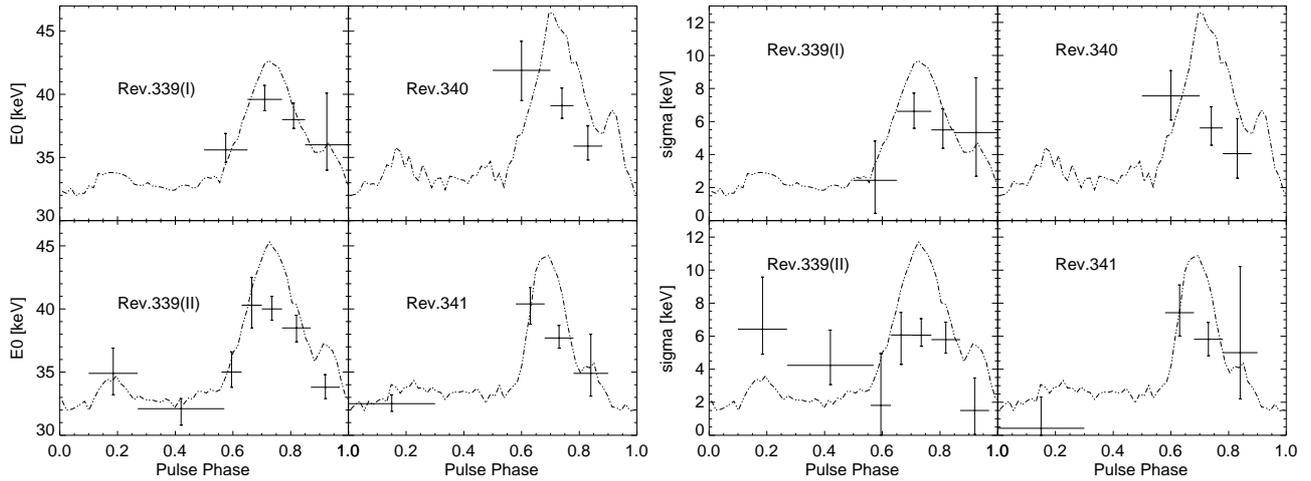}
\caption{\footnotesize {\em Left:} The position of the cyclotron line as a function of the pulse phase for four time intervals. Dotted line shows corresponding ISGRI pulse profiles in the range 20--100 keV. {\em Right:} The same for the width of the line ($\sigma$-parameter of the gaussian absorption line model). }
\label{prs}
\end{figure*}

In order to measure parameters of the cyclotron line as a function of the pulse phase we have split the data into four time intervals. During each interval the pulse profile does not change significantly and can be directly compared with the line parameters behavior. The result is presented in Fig.~\ref{prs}. The left panel shows the position of the cyclotron line with the pulse phase, the right panel shows the same for the width of the line. Dotted lines represent the corresponding ISGRI pulse profiles. Both parameters apparently show a positive correlation with the pulse-phase-resolved flux. 

\section{Simultaneous optical observations}

Optical observations of the source simultaneous with \hbox{INTEGRAL}
have been performed with the 1.25m ZTE telescope in SAI Crimea station
in July 28-30 2005 (MJD:53579-53581, $\phi_{\rm pre}\sim 0.12-0.18$). The
aim was to search for rapid variability with the period of X-ray
pulsations \citep[optical 1.24\,s pulsations have been reported by some authors, see e.g.][]{MiddleditchNelson76}
During our observations we failed to find any pulsations with
the amplitude higher than 0.3\% in the V band.

\section{Discussion and conclusions}

The shape of the X-ray pulse profiles of Her X-1 and their evolution with time is a subject of great discussions in the literature. There are two basic models explaining their behavior. Both of them assume a complicated structure of the emitting region on the surface of the neutron star. In the first model systematic variations of the pulse profile are due to a partial obscuration of the neutron star by the inner edge of the precessing accretion disk \citep[see e.g.][]{Scott00, BlumKraus00}. In the second model the evolution of pulse shapes is interpreted by free precession of the neutron star \citep[see e.g.][]{Truemper86,Ketsaris00}. 
To discriminate between these models one needs a larger "library" of high quality pulse profiles in different energy bands equally distributed in time. The INTEGRAL observations contributed substantially to this task. 

From the historical monitoring of the cyclotron line in the X-ray spectrum of Her X-1 there is evidence for a flux-related change of the cyclotron line energy. In particular, the RXTE observations in combination with data from INTEGRAL (this work) show an apparent correlation between the cyclotron line energy and the X-ray flux in the middle of the main-on \citep{Staubert06c,Staubert06d}. The correlation is positive, contrary to what is observed in some high luminosity transient pulsars \citep{Tsygankov06a,Tsygankov06b}.  

The pulse-phase-resolved analysis shows a significant variability of the line position with pulse phase, confirming previous results of \citet{Voges82} and \citet{Soong90b}. The INTEGRAL observations do not confirm the presence of a harmonic cyclotron line at $\sim75$ keV as reported by \citet{DiSalvo04}.

\section*{Acknowledgments}

The work was supported by the DFG grants Sta 173/31-2 and 
436 RUS 113/717/0-1 and the corresponding RBFR grants RFFI-NNIO-03-02-04003 and RFFI 06-02-16025, as well as DLR grant 50 0R 0302

\bibliographystyle{aa}
\bibliography{refs}

\end{document}